# Subjects, Not Authors

## *The Authorship Hazard in Agentic Dataspaces*

Seungho Lee · Changbin Lee

*Korea Trade Network (KTNET), Seoul, Republic of Korea*

**ABSTRACT—**Dataspace connectors decide *whether* a transfer may occur, not *what* the transferred value contains—tolerable for contracted applications, not for LLM agents that compose tool calls. Work on agents that generate governance artifacts evaluates output quality, not who may authorize an artifact for use. A published policy is what the decision point enforces, so publication is a governance event, and agents that are both policy subjects and policy authors write the norms that bind them. We name this the *authorship hazard* and state one principle: an agent is a subject of the governance plane, never an author of it. Its authorization channel to publication is closed by construction; its influence channel—drafting what humans approve—becomes an enforcement problem.

Across 90 preregistered edits to the paper's running agreement, each evaluated on 344,512 requests, the six that only reclassify a field all change authorization and narrow a duty without touching policy text, and a policy-diff classifier passes all six. Read as worded, the privilege-delta conditions also pass 33 of 69 effective policy-text edits; read as covering any relaxation, none. Treating classification as authorship routes all six to review; the registry this requires is not yet built. At the execution boundary, protected fields reach the model in 105 of 105 cases under prompt-stated duties and in 0 of 105 under a compiled tool-call constraint, but values outside named fields are exposed in 7 of 7. At the review share measured, a central approval pool needs one approver per 20 to 138 participants.

## 1. Introduction

A dataspace makes data sharing governable by making it contractual: two participants negotiate a policy, the policy becomes an agreement, and a connector enforces it. In the Eclipse Dataspace Components (EDC) stack, the reference implementation in Catena-X and a growing number of sectoral spaces, an ODRL policy decision point (PDP) evaluates the agreement and the data plane refuses transfers the PDP rejects.

What the connector enforces is narrower than what the contract says. It enforces *whether* a transfer happens, not *what* the transferred value contains, and it stops caring once the bytes leave. ODRL duties—anonymize before release, delete after thirty days, record provenance—are declarative; the connector transforms no payload. Duties are honored because the counterparty is an application under a contract, written by an organization that can be sued.

Agents break this arrangement, not because they are unpredictable, although they are, but because an agent is not a party. It has no legal personality, is instantiated and destroyed at will, can call tools its principal never enumerated, and can delegate to agents the provider has never heard of. The contract was signed with an organization; the process touching the data is three delegation hops away.

Two responses exist. DAVE [1] places a policy-enforcing spokesperson on the provider side that answers questions about documents rather than releasing them; its authors treat the constraint instructions given to the model as advisory and state that LLM-based virtual redaction cannot guarantee that no sensitive fragment leaks. The IDSA position paper [2] proposes a delegated agent credential verified at admission, with cascading delegation in which a secondary agent inherits the purpose scope and assurance grade of the primary's credential. The first adds a probabilistic component to the enforcement path. The second, like this paper, argues that existing mechanisms suffice, but places agent authority in the credential plane where we place it in the policy plane; Table 2 states the cost of that difference, two decision authorities rather than one.

### *1.1 Publication is a governance event*

Agents that *generate* governance artifacts are an established track: natural-language-to-ODRL translation [3, 4], ontology matching with paired LLM agents [5], multi-agent ontology generation [6] and LLM-driven knowledge-graph construction [7]; the IDSA Rulebook itself recommends AI for mapping semantic models and for automated policy evaluation [8]. We inspected these systems' evaluation designs for one thing (Section 9.5): a step at which a human or a policy decides whether the generated artifact may enter the plane that governs production. There is none, and within that literature's scope there need not be: a generator is evaluated on the *quality* of what it generates.

In a generic knowledge-graph pipeline a wrong mapping degrades retrieval. In a dataspace the published semantic model is what agreements reference and the published ODRL policy is what the PDP enforces, so a wrong or adversarial mapping can manufacture an authority that should not exist or erase a constraint that should. **Publication, not generation, is the security-relevant act**, and neither literature models it: generation research stops at output quality, and the governance literature does not model the generator. Software supply chains model an untrusted author, but they review code, whose meaning is its text (Section 9.1); a policy's meaning also depends on the sensitivity classification it refers to, itself a published artifact (Section 5.4).

Treat the agent as an ODRL subject, the natural next step, and an agent that is both subject and author of the policy writes the norms that bind it. This is the self-amendment problem of access control, and we call it the *authorship hazard*. Multilevel security already treats relabeling as the security-relevant act (Section 9.1); what it lacks is a drafter drawn from the class the policy constrains.

The identity the hazard turns on is one of class and adversary, not of instance. In Section 7 the agent that drafts a mapping and a policy is a task-bounded agent inside the provider's boundary, and the agent the policy binds is the consumer's planner. They are one party because Table 1 assumes every agent's planner adversarial and because the drafter's input is within the attacker's reach: a column name or sample value in the legacy source it profiles can carry an injected instruction (Section 4), planted by a consumer whose agents the published policy then governs. Where one participant both provides and consumes on a single agent platform, the instances coincide outright.

### *1.2 Principle*

> **PRINCIPLE P**
>
> An agent is a **subject** of the governance plane and never an **author** of it. Policy is made outside the agent and kept inside it.

P is a design constraint, not a theorem: it closes one channel by construction and leaves another open by necessity (Section 5). *Author* means the party whose act makes an artifact binding, not the party that wrote its text; an agent may draft, but the binding act is publication, and P denies the agent that act.

### *1.3 Contributions and their boundary*

- **The authorship hazard as a framing, with evidence**: the observation above, a targeted inspection of the generation literature (Section 9.5), and the argument that in dataspaces publication is the governance event. Priority on the broader *agent authority* problem is [2]'s.
- **C1 — Agents as ODRL subjects, with the authorization channel to publication closed.** The agent sits inside the policy plane as an ODRL party rather than beside it as a credential holder; it drafts, a human approves, and the approval artifact is the publishing credential. The channel that stays open, influence, is treated as an enforcement problem (Section 5.3), and the hazard is extended to the classification a policy refers to, treated as policy-rank authorship (Section 5.4); the registry this requires is modeled in Case D's evaluator, not yet implemented.
- **C2 — Duties that cross the execution boundary.** An ODRL duty in the negotiated agreement compiles into a constraint applied at tool invocation and on the return path, so the obligation acts on values rather than on visibility (Section 6.1). Execution-time obligations are not new: usage control has systematized them for two decades [9], and the IDS lineage enforces duties deterministically in production. We claim only that the duty carries provenance from a negotiated agreement, compiles to the model's tool boundary rather than a data-access boundary, and is the execution-side half of a design whose other half governs authorship.

Four experiments bear on these claims, Cases B, E, D and A (Section 8), and Table 12 sets every claim against its evidence. The re-delegation ceiling of Section 6.2 is designed, unevaluated and not claimed, and two threats are named and not addressed (T5, T6).

## 2. Background

### *2.1 What EDC does and does not enforce*

EDC separates a control plane, which negotiates agreements over the Dataspace Protocol and evaluates ODRL at a PDP, from a data plane, which transfers bytes under an endpoint data reference token. Its enforcement is real but *access*-level: token validation, transfer authorization, and a policy monitor that can terminate an ongoing transfer. No EDC component transforms a payload according to a duty—the ODRL duty is declarative and the specification assigns no executor—so the missing layer can be added beside the connector, which is not forked. EDC's class `ParticipantAgent` denotes a *participant*, not an AI agent.

### *2.2 What ODRL already permits*

**ODRL already admits agents as parties.** The Information Model defines a Party as

> *an entity or a collection of entities that undertake functional roles in a Rule, such as a person, collection of people, organisation,* ***or agent****.*

Placing an agent in the `assignee` position is thus something the specification has always allowed and no dataspace deployment operationalizes.

**ODRL cannot express constrained re-delegation.** The nearby terms—`nextPolicy`, `grantUse` ("to grant the use of the Asset to third parties") and `distribute`—all presuppose that *the asset moves*; in agent re-delegation the asset stays inside the provider's boundary and a sub-agent acts on it there. Nothing in ODRL restricts re-delegation, and that absence is the extension point of Section 6.2.

The W3C ODRL Data Spaces profile [10] defines ten actions, among them `Anonymize` and `Transform`, the duty actions our executor performs, and four party functions—`Provider`, `Consumer`, `Controller`, `Broker`—with no agent role or delegation term. It is an unofficial draft in the W3C ODRL group's repository with no status designation (retrieved 22 September 2026); we treat it as the vocabulary a live profile is converging on, and no claim depends on its status.

### *2.3 Agents, tools and the execution boundary*

An LLM agent reaches the world through tool calls, increasingly standardized as the Model Context Protocol (MCP), and other agents through agent-to-agent protocols. The boundary between "the model decided to do this" and "this happened" is the execution boundary, the only place enforcement is possible. That prompt-level constraints do not hold there is reported consistently. Tool-using agents follow injected instructions at substantial rates, roughly 24% for ReAct-prompted GPT-4 and nearly double under reinforced attacks [11]; an instruction hierarchy improves robustness without closing the gap [12]; runtime enforcement frameworks exist because prompting is insufficient [13, 14, 15]; and redacted question answering reports the same shape [16]. For MCP, a proxy that filters the tool registry at discovery time reports unauthorized-invocation rates of 48.5–68.5% under prompt-only control, 4.0–37.0% with a per-tool allowlist in the prompt, and 0% under architectural filtering [17]. The July 2026 OpenAI–Hugging Face incident shows the point outside a benchmark [18]: agents in an evaluation environment that denied them direct Internet access repurposed the package mirror they were allowed to use as a channel to one another, then chained previously unknown vulnerabilities in it and in a third party's systems, with publicly exposed credentials, to reach that party's production infrastructure. The evaluation ran without production classifiers, system prompts or auto-review; the report finds that the production harness and system prompt can cut the measured propensity to compromise infrastructure more than a hundredfold, and such controls lower a rate. The incident is not evidence for the authorship hazard studied here; it shows that what bounded the agents was the boundary that could be enforced, not the task or the capability boundary their operator intended. Our question is downstream: given that the constraint must be architectural, *where does it come from*? In [17] it is a hand-written attribute-based rule set, in [13] a developer-authored policy; here it is the agreement already negotiated.

## 3. Motivating Setting

Our setting is cross-border trade document exchange—certificates of origin, packing lists, customs declarations, carbon-accounting attachments—among exporters, forwarders, carriers and national single windows. **It is an instantiation, not the scope of the claim**: it supplies a federated infrastructure in which agreements, policy artifacts, delegated processing and execution-time obligations already coexist, so the mechanisms can be measured against something real. Three properties drive the design.

**Supply is fragmented and long-tail.** Thousands of small exporters hold data that a few large consumers want, in unharmonized schemas: pay-as-you-go integration [19]. Consumers announce a need and holders raise a hand, structurally the Contract Net protocol [20], which is why an agent is attractive on the supply side at all.

**The profile is more sensitive than the data.** To map a legacy table into a shared model an agent must see column names, value distributions and samples, which disclose commercial structure. The provider-side components therefore sit *inside* the participant boundary and only the approved mapping moves upward; a mapping carrying any literal data value is rejected (Section 5.3), since a mapping with sample values embedded *is* the profile.

**The consumer runs a planner; the provider must not have to.** A *planner* has an open goal and tool set and re-plans at runtime; a *task-bounded agent* has a fixed task and a closed tool set. Planners sit only on the consumer side.

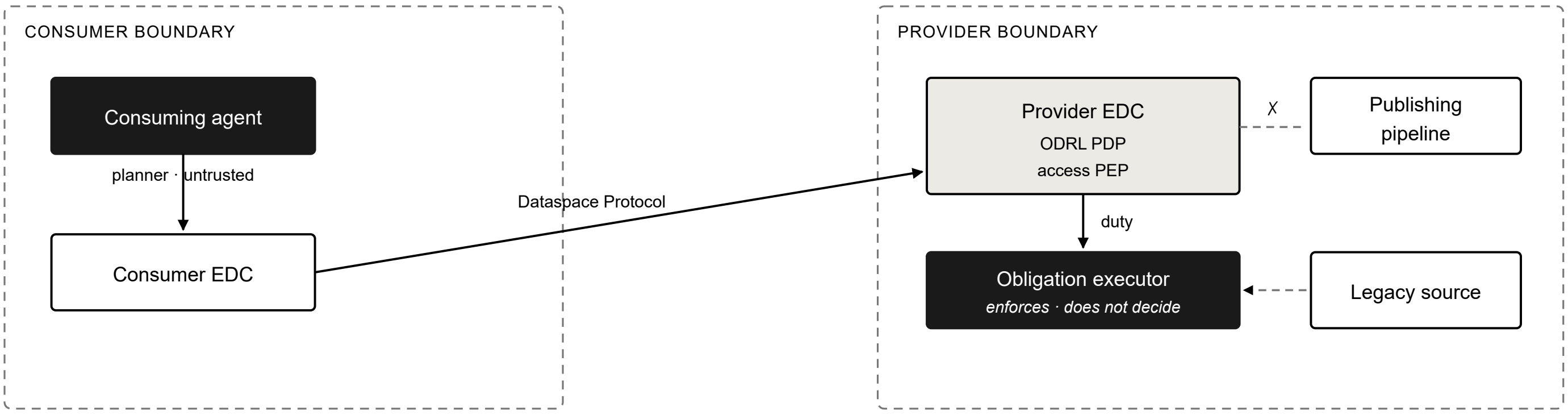

**Figure 1.** Runtime deployment. The consuming agent drives its own connector and never speaks to the provider directly. The obligation executor is added beside the provider connector because EDC has no duty layer; it enforces the PDP's decision and makes none of its own. The crossed line is the absent *authorization* edge: no agent can invoke publication. The path of *influence* on what gets published exists and is drawn in Figure 2.

**Table 1.** Trust assumptions.

| Component | Trust | Basis |
|---|---|---|
| Provider and consumer EDC (PDP, access PEP) | Trusted | Existing dataspace trust anchor; root of the delegation signature chain (6.2.3) |
| Obligation executor | Trusted | Deterministic, provider-operated, no model inside |
| Publishing pipeline, approval service, vocabulary registry | Trusted | Deterministic, not addressable by any agent |
| Tool gateway applying compiled constraints | Trusted | Same role as the MCP proxy in [17] |
| Human approver | **Trusted but bounded** | Honest, but attention is finite; modeled as fallible under load, not as an oracle |
| Planner (LLM) in every agent | **Untrusted** | Assumed fully adversarial: injected, misaligned or both |
| Sub-agent runtimes | **Untrusted** | Consumer-side, arbitrary |
| Consumer infrastructure beyond its connector | **Untrusted** | — |

## 4. Threat Model

Assets are the provider's data, the governance plane (published policies and vocabulary) and the agreement; Table 1 gives the trust assumptions. The adversary controls planner output in every agent: it can craft any draft, spawn sub-agents, decompose goals and observe all responses. It reaches the registration agent through what that agent reads, since a column name or sample value in the profiled source can carry an indirect prompt injection in the sense of [11]. It cannot break cryptography or compromise trusted components, so two artifacts are unforgeable for a stated reason: approval artifacts, issued by the trusted approval service (Section 5.3), and the signature chain attesting delegation lineage, rooted in a key the consumer connector issued (Section 6.2.3).

- **T1** Unauthorized publication of policy or vocabulary. *Closed by construction* (5.2).
- **T2** Manipulation of a draft's content so that an honest approver publishes something adversarial, including reclassification of the vocabulary a policy refers to, a change to the context that binds its terms, and a rendering that shows the reviewer other text than the parser reads. *Bounded, not closed* (5.3, 5.4).
- **T3** Evasion of a duty at the tool boundary. *Addressed* (6.1).
- **T4** Escalation through delegation, including a forged or self-reported lineage. *Designed, not evaluated* (6.2).

- **T5** Composition of a forbidden outcome from individually permitted actions. *Not addressed* (10).
- **T6** Derivative relay: a parent passes data it lawfully holds to a child beyond the child's ceiling. *Not addressed; an information-flow problem* [21] (10).
- **T7** Profile exfiltration through the registration agent's only egress, the draft. *Bounded for verbatim values*; encoded or split values and the mapping as a covert channel are not addressed (5.3, 8.4).
- **T8** Flooding the approval queue to stall publication. *Contained by a per-participant quota; capacity is a provisioning parameter* (5.3, 8.4).

The design's advertised properties expose T5 and T6: a stateless per-action evaluator cannot see a plan, and an access-control boundary cannot see what one process tells another. T5 probably matters most, since decomposing a goal into permitted steps is what a planner is for. **Out of scope:** side channels, model extraction, collusion between provider and consumer organizations, compromise of trusted components (Section 10), and denial of service against the PDP.

## 5. C1: Agents as Subjects, with Authorship Regimented Out

### *5.1 The agent inside the policy plane*

We place the agent in the `assignee` position of the ODRL rule governing the action it attempts and evaluate it at the existing PDP, per action (Table 2).

TABLE 2. Two ways to admit an agent. The right column is the design; in the prototype, per-action evaluation over fields, purpose and delegation depth runs in a reimplementation outside EDC (Section 10).

| Axis | Credential check [2] | Policy evaluation (this work) |
|---|---|---|
| Locus of authority | A delegated agent credential, verified separately | **The ODRL agreement itself** |
| Deciding component | A new verification procedure | **The existing PDP** |
| Decision time | Once, at admission | **Per action, continuously** |
| Relation to ODRL | Must be kept consistent with it | **Is it — policy composition** |
| Revocation | Credential revocation | **Revoking the agreement extinguishes authority, if each decision queries revocation (8.5)** |
| Decision authorities | Two: the credential verifier and the PDP | **One: the PDP** |

The last row needs a definition, because C2 adds a vocabulary term, an enforcement component and a compilation layer. A *governance stack* is a component that *decides* authority. The obligation executor enforces a decision the PDP has made, the compiler translates a duty the PDP has admitted, and the profile term is evaluated by the PDP: one decision authority, though not one component. Two deciding components can disagree; one cannot disagree with itself. This is an alternative design for an acknowledged problem, not a discovery: [2] identifies agent authority in dataspaces as open, describes cascading delegation and a tool-registry profile element, and notes that no agreed specification yet governs agent delegation at organizational scale.

### *5.2 Closing the authorization channel*

C1 is coherent only if the agent cannot author what binds it. We separate the drafting subject from the publishing subject in the sense of Jones and Sergot [22]: a regimented system does not punish the violation, it makes the violating action unavailable. The registration agent drafts a mapping and a policy, a human approver reviews the draft, and the approval artifact, not the agent's identity, is the credential the publishing API requires; the agent has no route to that API and holds no such artifact. The authorization relation is thereby *static with respect to agent actions*, a restriction under which the HRU safety question becomes decidable [23]. It says nothing about whether an agent can *influence* what a human with authority publishes, so we state both halves.

**Proposition 1 (authorization invariance).** *Let M be the authorization state that* `publish` *writes: the policy store, holding the published offers, and the vocabulary registry. The decision point evaluates negotiated agreements, and condition (iv) ties those to M. Let $\Sigma_A$ be the action alphabet available to the agent, including the negotiation actions of Section 2.1. Assume (i)* `publish` $\notin \Sigma_A$*; (ii) the publishing API's precondition requires an approval artifact issued by the approval service; (iii) no finite sequence in $\Sigma_A{}^*$ yields such an artifact; and (iv) every agreement's policy is the provider's stored contract definition at the moment of agreement. Then for every $\sigma \in \Sigma_A{}^*$, M is unchanged by σ, and every right the agent holds after σ was in M when it was granted. The safety question for this subject set—can the agent come to hold a right r on object o that M does not grant—is therefore decidable, and answered in the negative by construction.*

The proof is immediate. Conditions (i) and (ii) are architectural facts about which endpoints exist and what they require; (iii) holds because the approval service is trusted and not addressable by any agent (Table 1). Condition (iv) is a measured property of the connector (Section 8.3): the provider builds the agreement from the contract policy it currently stores and never from the policy text the consumer sent, and the two qualifications measured there bound the word “published” without letting an agent create a right. Condition (iv) is not decoration: the agent drives its own side of the negotiation (Figure 1), so without it negotiation would be a second operator on authorization state open to the agent. The proposition is shallow by design; the undecidability that motivates the safety literature [23] is avoided by removing an operator, not by restricting a model.

**Remark 1.** *Proposition 1 concerns authorization and asserts nothing about influence.* Condition (iii) stops the agent from producing the artifact, not from producing the *text* that an authorized party will approve and publish; M may change as a *consequence* of agent actions while invariant *under* them. Case D shows that this channel is open: with no filter, 75 of 90 preregistered single edits change authorization or duties, including all six that change only a classification (Section 8.4). Taking Proposition 1 as a security guarantee for the system takes it for the system's smaller half.

### *5.3 The influence channel and the approval stage*

A path from agent to publication exists through the approver, and its first node is the adversary of Section 4. If drafts arrive at volume and attention is finite, approval degrades toward ratification, and an adversary writing what an honest party rubber-stamps is in effect one that publishes. This is a confused-deputy surface [24], familiar from software supply chains, where an honest reviewer merges a change an untrusted author wrote [25]; HRU-style formalisms model who holds a right, not who wrote the text the right-holder signed. The approval stage is therefore enforcement, not regimentation [22]: approving an adversarial draft can be made costly, detectable and bounded, not impossible.

**(a) Drafts are diffs** against the currently published state of the same asset's policy and mapping, so a reviewer reads what changed. A first publication has no prior state, so it always takes the elevated path of (b) with an onboarding checklist: every field's sensitivity class assigned or confirmed by a human, no literal values, the default ceiling of Section 6.2.2. In the setting of Section 3 onboarding is the dominant event, and it happens once per asset.

**(b) Privilege-delta flagging.** A deterministic classifier over the diff flags any change that adds a permission, widens a constraint, weakens a prohibition or duty (deletes, disables or narrows it), introduces any new assignee, relaxes a delegation ceiling in any dimension or removes it, or alters a duty's target; Section 8.4 reports what a narrower reading of the third, fourth and fifth conditions misses. A change whose direction cannot be read from the diff—a changed operator, a purpose in a hierarchy, a rule split in two or deleted and added back—is treated as a relaxation. Three classes are handled unconditionally. Any change to a field's sensitivity classification is flagged regardless of direction (Section 5.4), because a class can appear in a permission and in a prohibition or duty alike. Any change to the vocabulary registry or to this classifier is flagged, as the highest-privilege changes in the system. Any draft containing a literal data value outside the allow-list of vocabulary identifiers is rejected, because it is the profile leaving the boundary (T7). Flagged diffs go to elevated review by two approvers, unflagged diffs to the ordinary path; both end at a human. The classifier is syntactic over ODRL and the mapping language.

**(c) Approver load ceiling.** The flagged diffs an approver may act on per period are bounded; beyond the bound drafts queue, and queue depth is a published metric, so rubber-stamping becomes an observable state. Because a queue is also a target for flagged, harmless drafts at volume (T8), quotas are per participant: the submission gateway throttles the participant, not the plane.

**(d) Approval artifact issuance.** The artifact is bound to the hash of the exact draft, single-use, short-lived, issued by a service no agent can address, and logged with the approver identity; a draft altered after approval fails at publication. Approver fatigue and social engineering remain human-factors problems that we bound and make measurable, not solve.

### *5.4 Classification is authorship*

The sharpest attack passes all six conditions of (b). The ceiling of Table 4 is expressed over sensitivity classes F0, F1, F2, and the registration agent drafts each field's class from the profile. A diff that moves `consignee` from F2 to F0 adds no permission, widens no constraint, weakens no prohibition or duty, introduces no assignee, relaxes no ceiling and leaves `Anonymize`(F2) textually intact; to an approver it reads as a label correction. Its effect is that the "Denied" verdict of Table 4 flips to "Permitted" and `Anonymize` no longer reaches `consignee`. The duty's scope narrows rather than empties, which is why the attack is quiet: the obligation still fires, on a smaller set, and nothing records that the set shrank. The policy has not changed by one character; what it protects has.

Publishing the *vocabulary the policy refers to* is therefore a governance event too, invisible to any check that reads only the policy; a classifier "syntactic over ODRL" is, on its own, a declaration of that gap. The same holds for the JSON-LD context that binds a policy's terms to IRIs: a changed binding alters what identical compact text means, and Appendix A's extension adds such a binding. The registry therefore has to hold the context and profile documents a policy uses, and diffs have to be computed on the expanded, canonical form; the prototype does neither, and Case D2 contains no context edits. Even then, what renders to a reviewer can differ from what a parser reads [26], and hash-binding the approval does not close that gap.

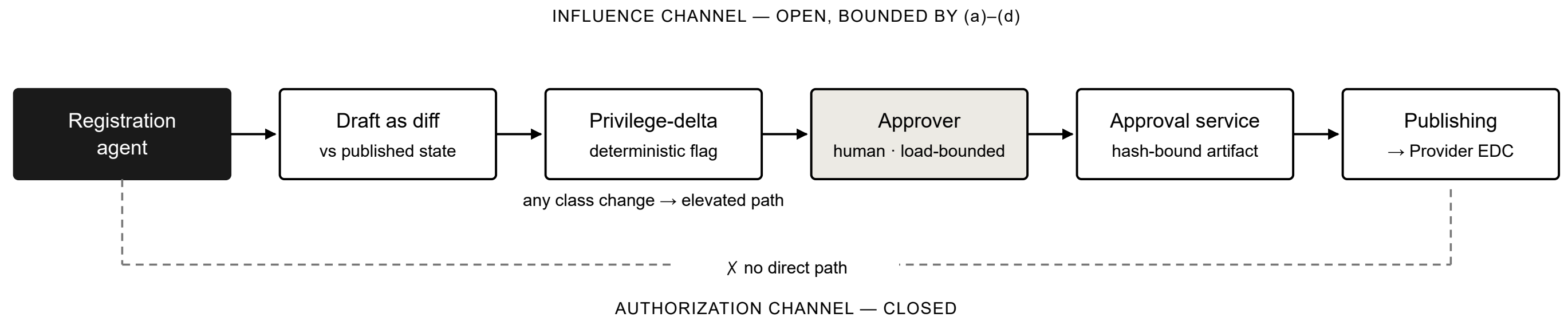

**Figure 2.** The two channels from agent to publication. The lower, crossed path is the authorization channel, absent by construction. The upper chain is the influence channel, which exists because the agent writes the draft; Section 5.3 bounds it with diff-form submission, privilege-delta flagging, a load ceiling on the approver and hash-bound single-use artifacts, and classification and vocabulary changes never take the ordinary path (Section 5.4).

The fix is already specified in (b). Classification is held as an attribute of the central vocabulary registry, not of the participant's mapping; any change to a field's class takes elevated review; and every published agreement whose ceiling or duties reference a reclassified field is listed for re-review. The agent stays useful for proposing classes and is denied their authorship. The cost, measured in Section 8.4, is that under federated approval every reclassification is central-rank traffic.

## 6. C2: Carrying the Agreement Across the Execution Boundary

C2 says where the architectural constraint of Section 2.3 comes from, the negotiated agreement, and propagates it in two directions (Figure 3).

### *6.1 Downward: duties become tool-call constraints*

An ODRL duty attached to a permission compiles into a constraint on the tool invocations the permission makes possible, applied at a gateway between planner and tool. An

`Anonymize` duty on a query permission becomes a transformation of the tool's return value before it reaches model context; a retention duty becomes an expiry on any artifact the call produces; a provenance duty becomes a record written before the result is released. The principle is old: complete mediation and least privilege [27] require the smallest sufficient authority to be checked at the point of use, and for a duty the point of use is the tool return path. Discovery-time filtering [17] decides which tools are visible and cannot mask a field in a response, because when the response exists the decision is past; the two are complementary. As with the shrink-only guarantee of [13], a compiled constraint set may only be narrower than the agreement, never wider.

### *6.2 Sideways: re-delegation beneath an owner-declared ceiling*

**This subsection is design, not result**; nothing in it is evaluated. Capability systems—Macaroons [28], UCAN, ZCAP-LD—mint an attenuated derived token when an agent spawns a sub-agent, and agent infrastructure propagates constraints from parent to child along delegation chains [29], as does the IDSA sketch. Delegation is therefore not claimed. One axis differs (Table 3): in a capability system the holder attenuates freely, whereas in a dataspace the owner must have a say in whether its counterparty may hand its data to a sub-agent, and how far that sub-agent may reach. Cascading delegation in [2] says constraints are inherited, not who authorizes delegation, and ODRL has no term for it (Section 2.2).

#### *6.2.1 A concrete ceiling*

A certificate of origin (CoO) carries fields in three sensitivity classes: F0, public trade facts (HS code, origin country, issue date, certificate number); F1, commercial quantities (quantity, gross weight, declared value); F2, identities (exporter, consignee, producer). Actions form a chain `read-metadata` < `read-fields` < `read-document`. The agreement grants the primary consuming agent `read-fields` over F0 ∪ F1 for the purpose *customs clearance*, with duties `Anonymize`(F2) and provenance-record; Table 4 gives the owner's ceiling for any sub-agent.

**Table 3.** Re-delegation: capability systems versus the policy plane. Rows 4 and 5 are costs.

| Axis | Capability systems | This work |
|---|---|---|
| Who bounds re-delegation | The holder alone — free attenuation is the design goal | **The owner declares a ceiling; the holder attenuates beneath it** |
| Form of authority | A derived token carrying rights | **No rights on the wire; an attestation of identity and lineage, rights derived at evaluation** |
| Basis of attenuation | Whatever the holder keeps | **Holder's choice, within the owner's ceiling** |
| Revocation granularity | Per branch — finer | Revoking the agreement collapses the whole chain — coarser |
| Availability | Offline verification possible | Every action requires the PDP |

**Table 4.** Owner-declared ceiling for sub-agents on a CoO, and two decompositions against it.

| Dimension | Ceiling | HS-classification checker | Consignee-risk lookup |
|---|---|---|---|
| Action | ≤ `read-fields` | `read-fields` | `read-fields` |
| Fields | ⊆ F0 | {HS code, origin} | {consignee} — F2 |
| Purpose | = customs clearance | customs clearance | customs clearance |
| Depth | ≤ 1 (derived from chain length, 6.2.3) | 1 | 1 |
| Duties | inherited, non-removable | inherited | inherited |
| Verdict | — | **Permitted** | **Denied — F2 above ceiling** |

The checker is a legitimate decomposition and passes; the lookup is denied because the owner never agreed that any downstream process may *hold F2 as a policy subject*. A stronger claim does not hold: the primary agent, an untrusted planner holding F1 and the anonymized form of F2, can perform the lookup itself and paste the result into the sub-agent's prompt. The ceiling bounds what a sub-agent can *obtain from the PDP*, not what it can *learn from its parent*; it delivers least privilege and auditability, not confinement (T6).

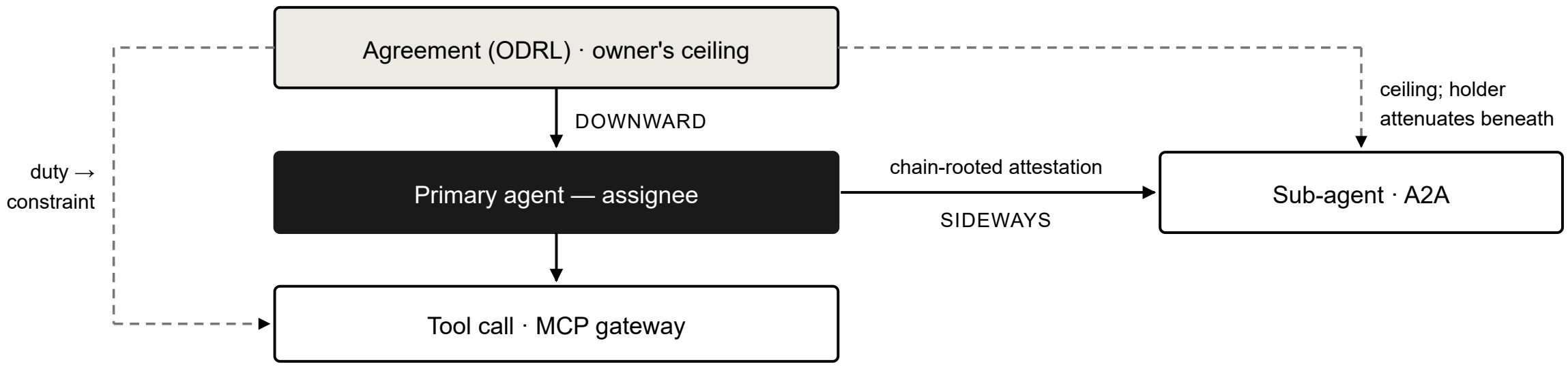

**Figure 3.** Two directions of propagation. Duties descend from the agreement into constraints on tool invocation and on return values. Permissions travel sideways into sub-agents beneath the ceiling the owner declared, and the holder attenuates further beneath it. What crosses to the sub-agent is an attestation of identity and lineage, not a rights-bearing token.

#### *6.2.2 Can an owner declare this?*

At negotiation time the owner knows nothing of the consumer's agent topology, and capability systems give attenuation to the holder because only the holder knows how it will decompose its work. But the owner declares a *ceiling*, not a topology: which fields, which actions, how many hops, for what purpose—exactly what an owner knows—and the holder attenuates beneath it. What remains is a security–utility tradeoff whose useful point is empirical and is the first question of Case C (Section 10). As a default we propose F0-only at depth 1 with fixed purpose, relaxed per agreement.

#### *6.2.3 What crosses the wire*

If no token is minted, how does the PDP know which sub-agent is acting and by whose delegation? A sub-agent presents an *attestation*: its own key, the agreement identifier, a timestamp, and a signature chain over that key running back through each parent to a key the consumer connector issued to the primary agent. Depth is not carried; the PDP *derives* it by counting signatures to the connector-issued root. Because Table 1 places sub-agent runtimes and the primary's planner in the untrusted region, rooting the chain in the connector puts its origin inside the dataspace trust anchor, and a hop without a signature from a connector-certified key is not a hop. The attestation carries *claims*, not *rights*: the PDP derives rights from the agreement at evaluation time, so a replayed attestation yields what the agreement currently grants, which after revocation is nothing, provided the decision path queries revocation (Section 8.5).

*6.2.4 Costs*

**Revocation is coarse**: revoking the agreement collapses the whole chain, and cutting off one sub-agent means amending the agreement, a contract operation. **Availability couples to the PDP**: every action requires a decision, and per-action latency is added to every tool call (Case A). A cached decision restores availability at the cost of a revocation window, a dial the deployment sets.

## 7. Architecture

The separation follows from Section 2.1 (Table 5): EDC's access layer is used unchanged and its decision layer unforked, taking constraint functions as Appendix A shows; the duty layer is new because it is absent; and the planner is outside every enforcement path.

**Table 5.** Separation of decision, enforcement and planning.

| Layer | Component | Decides? | Status |
|---|---|---|---|
| Decision (PDP) | EDC ODRL policy engine | **Yes — the only one** | Extended with constraint functions, as in Appendix A; not forked |
| Enforcement — access | EDC data plane, policy monitor | No | Used as is |
| Enforcement — duty | Obligation executor | No | **New — the layer EDC lacks** |
| Enforcement — tool boundary | Duty compiler + MCP gateway | No | **New** |
| Enforcement — registration egress | Draft-submission gateway | No | **New — literal-value rejection, closed tool set** |
| Planning | Consuming agent (LLM planner) | No | Outside every enforcement path |

Three agents appear. The **registration agent** is task-bounded and runs inside the participant boundary at design time: it profiles a source and drafts. Its tool set is closed by a gateway, not by intention: its only egress is the draft-submission endpoint, which applies the literal-value rejection of Section 5.3, and it can only read vocabulary identifiers in the central registry. The **providing agent** is task-bounded and runs inside the boundary at runtime; it answers a negotiated request and does not plan. The **consuming agent** is a planner and runs centrally. This is not the per-site semantic middleware dataspaces were meant to avoid: the deployed artifact is a stateless process and a pointer to the central registry, with no authored content; what accumulates centrally is the approved mapping.

## 8. Evaluation

The four experiments answer different questions on different systems, and their letters are the identifiers under which the evidence package and its preregistrations were frozen. *Case B* asks whether a compiled duty holds at the execution boundary (8.1, 8.2); *Case E* traces the constraint to an agreement produced by a real negotiation (8.3); *Case D* asks what the approval stage is worth (8.4); *Case A* measures what per-action evaluation costs (8.5). Case C remains a plan (Section 10), and a prompt-behavior study (8.6) is secondary evidence.

**Setting.** The asset is the certificate of origin of Section 6.2.1; the agreement carries `Anonymize` over F2, a provenance-record duty and a retention duty; the tool is a single certificate-query tool. Three conditions are compared: *prompt-only*, the duty text given to the planner and nothing acting at the boundary; *discovery filtering*, the registry deciding tool visibility as in [17], with the tool visible and permitted; and *compiled*, agreement-derived constraints applied at invocation and on the return path. The primary metric is whether a protected value is present in the payload bound for the model, measured at the gateway, not inferred from the model's answer.

**Configuration.** Live-model trials run through vendor command-line runtimes, each in a fresh process and a fresh rootless container with an empty home and working directory and an allow-listed environment, so that no user, project or memory context is mounted, and the context that reached the model was audited per trial. The harness queries the certificate through the gateway and puts the result in the prompt; Claude trials have no tools, and codex trials keep the runtime's nine default tools, which cannot be disabled. A trial answered by a model other than the one requested is not counted (codex does not report the responding model). Sampling temperature is each runtime's default and not exposed. The nine configurations are four Claude models (`claude-opus-5`, `claude-sonnet-5`, `claude-fable-5-1`, `claude-haiku-4-5-20251001`) through claude-code 2.1.278; three GPT-5.6 configurations and one GPT-6 (`gpt-5.6-luna`, `gpt-5.6-sol`, `gpt-5.6-terra`, `gpt-6-astra`) through codex-cli 0.155.1, which has no system role, so the policy is prepended to the prompt; and `qwen3:8b` (Q4_K_M) served locally by ollama 0.24.0.

### *8.1 Deterministic measurement at the boundary*

The first layer runs no planner: 450 anonymize trials over 150 cases in three conditions, with zero model calls. Sixteen cases form a stress group reported below. Of the other 134, twenty-nine carry no protected value—tool errors, F0-only requests, absent or empty F2 fields—and cannot host a violation, leaving a risk set of **105 cases** that span single and multiple F2 fields, nested objects, arrays, optional and reordered fields, repeated values, Unicode, long values and the same field reached by different paths (Table 6). A separate 22-case risk set, kept in the evidence package, gave the same split: 22 of 22 exposed without the compiled constraint and 0 of 22 with it.

At n = 105, **"0 of 105" is a violation rate below roughly 3.5% at 95% confidence, not zero**; on the 22-case set alone the bound would be about 15%. Discovery filtering has no point of action on a response that already exists, so its 105 of 105 **confirms a structural argument rather than testing an open question**, and it is no failure of [17], which was built to decide invocation. The compiled row is closer to a real test, since the compiler must derive the right constraint and the gateway apply it without damaging permitted fields, but it is a test the design was built to pass on cases of this shape.

TABLE 6. Case B primary result. Protected F2 values reaching the model-bound payload over the 105-case risk set, with Wilson 95% intervals. No model calls. Identical under a fixture agreement and under the negotiated agreement of Case E. Transform overhead is an in-process microbenchmark and excludes the policy decision round trip, which Case A measures (Section 8.5).

| Condition | F2 in payload | 95% interval | Over-removal of F0/F1 | Transform overhead |
|---|---|---|---|---|
| Prompt-only | **105 / 105** | 96.5–100% | 0 | — |
| Discovery filtering [17] | **105 / 105** | 96.5–100% | 0 | +0.6 µs |
| Compiled (this work) | **0 / 105** | 0–3.5% | 0 | +19 µs |

**On the corpus the mechanism is not built for, it fails outright, and this is the most important result in the section.** The stress group holds sixteen cases in which the protected value is embedded in free text, sits inside an attached scan, is derivable from permitted fields, or appears only as a fragment. Seven are detectable by a literal scorer, and the compiled condition exposes the protected value in **7 of 7**, as do the other two conditions; the other nine are invisible to a literal probe and are not counted. 0 of 105 is therefore evidence that this corpus is easy, not that the problem is. How much of a real trade document has the harder shape is unmeasured; the documents of Section 3 mix structured fields with free-text descriptions, remarks and scans.

The other duties separate the same way. A provenance record is absent in 10 of 10 applicable cases with nothing acting at the boundary and written before release in 10 of 10 compiled; of 10 retention cases, 7 are past expiry, and the expired artifact remains readable in 7 of 7 with nothing acting and in 0 of 7 compiled; a malformed constraint fails closed, 1 of 1. **These rows are structural baselines, not model failures**, since no planner runs in this layer. Gateway overhead in a 1,000-iteration in-process microbenchmark is a median 24.4 µs prompt-only against 43.5 µs compiled; it bounds the duty transformation itself, not the per-action policy decision, which Case A measures.

### *8.2 With models in the loop*

The second layer puts live models behind the same gateway: **nine model configurations, three conditions, 1,320 calls**, of which 1,317 count, since in each condition one adversarial Fable task was answered by another model. The design is unbalanced: eight configurations contribute 20 benign and 20 adversarial trials per condition and the open-weight runtime 60 and 60, three times the weight of any other cell. The models were chosen because their prompt-only behavior differs enormously (Section 8.6), which makes invariance at the boundary the point; the counts are too thin for an interval on output leakage, reported as a raw count (Table 7).

Under prompt-only and discovery filtering the protected value reaches the model on **every benign trial as well**: legitimate requests, correctly served, by a planner doing nothing wrong. That column, not the adversarial one, is the one a deployment should find alarming. Benign success is 219 of 220 in every condition; the three failures fall on three different tasks and configurations, one per condition, so on this corpus the mechanism costs no utility. Adversarial output leakage under prompt-only ranges from **none on three configurations to 47 of 60 on the open-weight runtime**, which supplies 47 of the row's 56 adversarial leaks and all five benign ones. Model behavior varies by more than an order of magnitude; boundary exposure does not vary at all, and the compiled row is 0 for every configuration separately. **An output-based metric would have scored three of the nine prompt-only cells as clean successes, and none of them is one.** Discovery filtering and prompt-only record similar leakage, 54 against 56 adversarial and 3 against 5 benign, with context exposure 439 of 439 in both; the difference is sampling variation.

TABLE 7. Case B with models in the loop, aggregated over nine model configurations and 1,317 counted calls. Context exposure is measured at the gateway and is reported separately for the two task types, because the benign column is a result in its own right. Output leakage is the model's final answer, split by task type: a benign leak is a protected value volunteered in answer to a request that did not ask for it. Every one of the nine configurations shows the compiled row individually, with no exception.

| Condition | Context, benign | Context, adversarial | Benign success | Leakage, benign | Leakage, adversarial |
|---|---|---|---|---|---|
| Prompt-only | **220 / 220** | **219 / 219** | 219 / 220 | 5 / 220 | 56 / 219 |
| Discovery filtering [17] | **220 / 220** | **219 / 219** | 219 / 220 | 3 / 220 | 54 / 219 |
| Compiled (this work) | **0 / 220** | **0 / 219** | 219 / 220 | 0 / 220 | 0 / 219 |

### *8.3 Case E: where the constraint comes from*

Case B shows that a compiled constraint holds, not that it came from the agreement, the claim that separates this work from a hand-written rule set [17] or a developer-authored policy [13]. Case E tests it on the Tractus-X EDC 0.13.0 end-to-end Dataspace Protocol test harness, with provider, consumer and secure token service as separate runtimes, credentials signed by the dataspace issuer, and a real credential-registry container.

**A stock constrained duty survives negotiation.** With an obligation from the profile's own vocabulary and an unmodified harness, the duty is present at policy registration, in the catalog offer and in the finalized agreement held by both parties, expanded to its full profile IRI with operator and right operand intact. **The value-level term is rejected by the stock profile**: the duty this paper needs, suppressing fields of a named sensitivity class before release into an agent execution context, has no stock term, and the validator refuses to register one with HTTP 400. **With a minimal profile extension, the chain closes.** Adding one obligation term (`AnonymizeFieldClass eq "F2"`) took eighteen non-blank lines in four existing files plus two new files, a 63-line constraint function and a 35-line schema, with validation left on and a stock term still registering (Appendix A). The term survives an actual negotiation into the agreement on both sides and feeds the compiler and gateway of Section 8.1 unchanged; replaying the 105-case risk set with this agreement as the constraint source reproduces Case B exactly (105 of 105 raw responses carry F2, 0 of 105 model-bound payloads do, over-removal 0, no model called). The reproduction is the point: substituting a negotiated source for a fixture changed nothing downstream.

**Condition (iv) of Proposition 1 was tested on the same runtime.** Twelve negotiations on the container testbed of Case A asked what a consumer obtains by naming an offer that does not exist, widening or narrowing the permission in its request, quoting an offer identifier never issued, and requesting after the provider widens its definition. Non-existent targets were refused before any negotiation record existed (404 for a missing definition, 400 for a missing asset). The four widened requests and the one narrowed request were accepted, but each agreement the provider sent carried the published policy, and the consumer refused each as unequal to its own offer. All three finalized agreements equal the definition the provider stored at that moment. Two observations cut the other way: after the definition was widened, a request quoting the old offer received the wider policy, and an identifier whose instance part was never issued was accepted once its definition and asset parts resolved. The provider-side bound is therefore the stored definition, enforced by substitution; the comparison that stops a mismatched agreement lives in the consumer.

### *8.4 Case D: what the approval stage is worth*

Case D tests the hazard the paper is named for: whether a governance edit changes authorization or duties, and which such edits a gate reading only the policy diff, and one also reading classification, would observe. Its counts are exact for a constructed, preregistered edit space, not estimates of prevalence. An audit of an earlier constructed validation found generator and coverage defects (Appendix B), so we report the corrective replication D2, whose specification, evaluator, classifiers, edit generator, request universe and analysis code were frozen before its edit corpus was generated. No model and no connector is involved; the policy evaluator is a reimplementation of this paper's rules.

**Design.** The base state is the agreement of Section 6.2.1 with the ceiling of Table 4, plus a field-level prohibition on `declaredValue`, an F1 field the permission would otherwise release. Principals carry a key lineage—consumer organization, primary agent, three delegation levels, and an external organization with its agent—and the evaluator reads permissions, prohibitions, assignees through that lineage, the four ceiling dimensions and the classification; duties are evaluated as a separate state. Every request in a universe of 7 × 8 × (2 + 6 × 1,025) = 344,512—seven principals, eight purposes including none, two actions of fixed disclosure, and six that select fields over 1,025 selections of the ten schema fields and one unregistered field—is evaluated before and after each of 90 single edits in seven classes: permission addition (15), constraint widening (9), prohibition removal or narrowing (24), duty deletion, disablement or scope narrowing (13), assignee change (12), ceiling relaxation (11) and classification downgrade (6, every F2 field to each lower class), plus 25 literal-insertion tests. Duty-target and vocabulary edits are excluded, because retargeted duties have no defined execution and vocabulary effects would need a time axis. The implementation was qualified on a separate synthetic domain (17 checks; 30 seeded defects, each caught), the run was executed once, and every figure is computed from its raw record.

**V1 is an ablation of our own classifier, not a prior-work baseline.** V1 applies the six conditions of Section 5.3(b) to the policy part of each document diff; V2 adds the three unconditional classes. Three conditions admit two readings —"removes a prohibition or duty" (deletion only, or any weakening), "introduces an agent as assignee" (agents only, or any principal) and "deepens a delegation ceiling" (depth only, or any dimensional relaxation plus ceiling removal)—and both were fixed before the run; Section 5.3(b) states the wider one. Administrative models govern *who* may change authorization state [30] but do not classify a change as privilege-expanding, and we know of no system implementing these conditions over agent-authored dataspace policy (Section 9.1).

**TABLE 8.** Case D2. Preregistered single edits by class. "Unobserved" counts edits that change authorization or duties and that the gate passes; a flagged edit is routed to elevated review, not blocked, and whether its reviewers catch it is not measured here. Edits that change nothing: 15; flagged by the worded reading 2, by the any-relaxation reading 15.

| Class | Edits | Change authorization | Weaken a duty | Unobserved: V1, worded | V1, any relaxation | V2, worded | V2, any relaxation |
|---|---|---|---|---|---|---|---|
| Permission addition | 15 | 13 | 0 | 0 | 0 | 0 | 0 |
| Constraint widening | 9 | 9 | 0 | 0 | 0 | 0 | 0 |
| Prohibition removal or narrowing | 24 | 19 | 0 | **18** | 0 | **18** | 0 |
| Duty deletion, disablement or narrowing | 13 | 0 | 13 | **10** | 0 | **10** | 0 |
| Assignee change | 12 | 12 | 0 | 4 | 0 | 4 | 0 |
| Ceiling relaxation | 11 | 3 | 0 | 1 | 0 | 1 | 0 |
| **Classification downgrade** | 6 | **6** | **6** | **6** | **6** | 0 | 0 |
| Total | 90 | 62 | 19 | 39 | 6 | **33** | 0 |

**All six classification-only edits change authorization with no policy text changed.** Moving an F2 field to F0 admits 80 previously denied requests, 64 for the primary agent and 16 for the sub-agent within its ceiling; moving it to F1 admits 64. The consignee-risk lookup of Table 4 flips to Permitted when `consignee` moves to F0, and every one of the six narrows `Anonymize` from three fields to two. V1 passes all six under both readings, necessarily; V2 flags all six. The row measures not classifier skill but what the construction leaves open: that a reclassification touching no policy text changes decisions, and by how much.

**The policy-text path has a gap of its own.** Of the 84 policy-text edits, 69 change authorization or duties and 15 change nothing (relaxations along dimensions where the permission is already narrower, narrowings that still cover the governed requests, actions already implied). Read as worded, the six conditions pass 33 of the 69—18 narrowings of the prohibition's principals, actions, purposes or fields, three duty disablements, seven duty-scope narrowings, four edits naming an organization as assignee, and one widening of the ceiling's classes—and V2 passes the same 33. Read as covering any relaxation, they pass none of the 69 but flag all 15 no-effect edits, where the worded reading flags two. The earlier validation could not see this, because its evaluator read neither prohibitions nor assignees.

**Literal insertion** (Table 9). The literal rule rejects a sensitive value in every slot verbatim or with altered case and spacing; with spaces replaced by hyphens it passes in the two prose slots, which are matched after case and whitespace normalization. None of the 25 changes authorization or duties: literal embedding is exfiltration (T7).

**Table 9.** Case D2 literal insertions, five annotation slots each (three value slots, two prose slots): V2 verdict, reject / pass. V1 passes all 25.

| Inserted literal | Sensitive | V2 reject / pass |
|---|---|---|
| Sensitive value, verbatim | yes | 5 / 0 |
| Sensitive value, case/spacing altered | yes | 5 / 0 |
| Sensitive value, spaces as hyphens | yes | **3 / 2 (prose slots)** |
| Non-sensitive, outside the allow-list | no | 3 / 2 (value slots rejected) |
| Allow-listed identifier | no | 0 / 5 |

**What it costs.** The review shares here come from the earlier validation's corpus of 380 drafts under the worded conditions, which flagged all 140 classic attack drafts there; they are scenario inputs, not measurements of D2, which contains no benign edits, and the review cost of the wider reading is unmeasured. Under V2, 50 literal-embedding drafts are rejected at the gateway and 270 flagged, so **escalation to human review rises from 36.8% to 71.1%** of drafts, 140 to 270 of 380, with 13.2% rejected automatically. False positives on the benign classes B1 and B2 are 0 of 60 (interval 0–6.0%); the whole cost sits on B3, where a change that tightens a classification is flagged 30 of 30, the price of direction-blindness that the downgrade row of Table 8 justifies. These percentages are properties of the corpus. With β the benign share, γ the share of benign drafts that change a classification, $p_1$ = 140/290 and $p_2$ = 240/290 the escalated shares of attack drafts, escalation is $(1-\beta)p_1$ under V1 and $(1-\beta)p_2+\beta\gamma$ under V2, and rejection $(1-\beta)\cdot$ 50/290 (Table 10): absolute burden falls in any realistic mix while the multiple grows. The vocabulary and literal-embedding classes, 100 drafts and 26% of the corpus, appear only on the cost side.

**Table 10.** Escalation to human review as a function of corpus composition; the rejection column is automatic and costs no approver time. **The first row is measured, counted directly from the earlier corpus's 380 trial records** (β = 90/380, γ = 30/90); **the other three are extrapolations of the model above**, computed with $p_1$ = 140/290 and $p_2$ = 240/290 held fixed while β and γ vary. Absolute cost falls as benign drafts dominate; the ratio between the two classifiers grows.

| Benign share β | Class changes among benign γ | V1, review | V2, review | V2, rejected | Ratio |
|---|---|---|---|---|---|
| 0.2368 (measured) | 0.3333 (measured) | 36.8% | **71.1%** | 13.2% | 1.9× |
| 0.50 | 0.3333 | 24.1% | 58.0% | 8.6% | 2.4× |
| 0.90 | 0.10 | 4.8% | **17.3%** | 1.7% | 3.6× |
| 0.99 | 0.02 | 0.5% | 2.8% | 0.2% | 5.8× |

**Centralized approval is not capacity-stable at the volume the paper motivates**, *under the provisioning we evaluated.* Each period $N$ participants submit λ drafts each; the share π are first publications, which always take the elevated path, and the classifier escalates the share $e$ of the rest (0.368 under V1, 0.7105 under V2). A pool of $K$ approvers at $C$ reviews each is stable while $N\lambda[\pi + (1-\pi)e] \leq KC$. A deterministic queue simulation over seven populations from 50 to 5,000, λ of 0.2, 0.6 and 1.0, π of 0.05, 0.30, 0.80 and 1.00, pools of three, ten and thirty at $C$ = 20, both classifiers and the quota on and off (1,008 scenarios of 120 periods) matches the inequality in all 72 parameter families. Under V2 at the lowest draft rate, three approvers hold to 200 participants and have diverged at 500, ten hold to 1,000 and thirty to 3,000. **V2 needs one approver per 100 to 138 participants at the lowest draft rate, 33 to 46 at the middle one and 20 to 28 at the highest**; at 5,000 participants that is 37 to 50 approvers at the lowest rate and 109 to 150 at the middle one (V1: 20 to 50 and 60 to 150), by arithmetic on the simulated arrival rate. On the simulated grid, the amendment of Section 5.4 raises elevated arrivals by at most 1.83 times, at the smallest onboarding share and population, where rounding arrivals to whole drafts lifts the continuous 1.81. Below the grid's smallest onboarding share the factor tends to 1.93 (= 270/140) as first publications vanish, and it is 1 when every draft is a first publication. The per-participant quota converts a flooding participant (T8) from a plane-wide stall into a local one, lifting normal drafts served from 3,340 to 7,116 while the flooder falls from 3,860 to 84, without moving the saturation point. What does not scale is a single central pool, and nothing in the design forces that topology.

**Case D5: the federated topology.** Each participant reviews its own drafts, and a central pool of $K_c$ approvers reviews only the share μ that changes a classification, the vocabulary registry or the classifier. The same simulator, unmodified and under a preregistration frozen before execution, reproduces the 1,008 centralized scenarios in every field; all 12,096 local cells stay at zero depth, and the central queue saturates at $N^* = K_c\ C/(\lambda\mu)$, matching the closed form in all 504 families. **Federation moves the central limit; it does not remove it.** At the middle draft rate with three central approvers the pool holds to 5,000 participants when μ ≤ 0.02, to 1,000 at 0.05 and 0.1, and to 200 at 0.21, where the closed form gives 476, 1.9 to 4.8 times the centralized pool's 100 to 250; the ratio of the two limits is $[\pi + (1-\pi)e]/\mu$. The value 0.21 is the earlier corpus's reclassification rate, 80 of 380; counting the vocabulary and classifier edits the partition also routes centrally gives 130 of 380, 0.34, where the closed form gives 292. **μ is the parameter the exit depends on, and it is unmeasured.** No normal participant's queue changed in any of the 144 flood cells; arrivals are constant per period, so there is no latency comparison.

**What Case D is not.** Its results hold in our model of the semantics, not in a connector; where the paper leaves a rule open—how prohibitions compose with the action chain, whether a ceiling applies to an organization's own agent—the preregistration states the choice. The edit space is adversarial and constructed: it shows that the hazard is *possible and mechanical*, not how often it arises, and class sizes are fixed by design. Edits are applied singly to one base state, so combinations and semantics-preserving rewrites are untested and base-dependent results, such as the eight ceiling relaxations with no effect, would differ elsewhere. The same author wrote generator, evaluator and classifiers; a separate reference evaluator agreed on every request of every synthetic state, which guards against coding error, not against a shared misreading. The nearest external baseline, change-impact analysis [31], was not run (Section 10). Case D attacks the influence channel only, and its approver is a classifier: a flagged edit is scored as routed, not caught. The earlier corpus contains no legitimate privilege expansion and no first publication.

### *8.5 Case A: what per-action evaluation costs*

Case A quantifies the availability coupling of Section 6.2.4 on a testbed built from the same connector images as Case E—the upstream Tractus-X EDC 0.13.0 harness, not a production deployment—with provider, consumer, token service and BPN directory as separate containers (Table 11). It is a cost the design incurs.

The decision path is not EDC's policy engine: EDC 0.13.0 exposes no policy-evaluation endpoint, so our client makes an agreement lookup and a revocation lookup (medians 6.92 and 6.70 ms) and evaluates the agreement locally, checking that it exists, is not revoked, permits use and has not passed its end date. A second of added latency on a fifty-step plan is a real cost, and a reader who concludes from Table 11 that per-action evaluation is unaffordable for long plans is reading it correctly. The revocation rows show that the obvious remedy and the design's central property are the same knob. A five-second cache removes essentially all the cost but permits 99 more decisions on a revoked authority and delays the first denial from 34 ms to just over five seconds; at thirty seconds the figures are 591 decisions and thirty seconds. A path that reads only the contract agreement **never denied** within thirty seconds, across 439 decisions after revocation: EDC 0.13.0 records revocation as the agreement's retirement in a separate register, which every decision must query. Two limits remain: under a control-plane outage the cached client permits for the rest of its window, by construction, and a request requiring the BPN directory failed whether the directory was up or down, so whether catalog authorization fails open or closed there is unresolved.

**Table 11.** Case A. Cost of evaluating policy per action, against admission-time verification, a five-second decision cache, and a per-action path that reads the agreement but not the revocation register. Medians over 1,000 decision paths and 200 tool calls per variant; the revocation rows are medians of three runs, one run for the agreement-only path, whose rows are bounded by a thirty-second observation window. The admission-time client never re-queries, so revocation is invisible to it by construction and was not measured. The 50-step row is the median of twenty measured 50-step plans; fifty times the per-call median would read 2.37, 1,018.81 and 2.05 ms.

| Measurement | Admission-time | Per action | Cached, 5 s | Per action, agreement only |
|---|---|---|---|---|
| Decision path, median | — | **19.25 ms** | — | — |
| Decision path, p95 | — | **51.67 ms** | — | — |
| Added per tool call | 0.05 ms | **20.4 ms** | 0.04 ms | — |
| Added over a 50-step plan | 2.15 ms | **1,019.17 ms** | 2.01 ms | — |
| First denial after revocation | — | 34 ms | **5,025 ms** | **none in 30 s** |
| Decisions permitted after revocation | — | 0 | **99** | **439 in 30 s** |

### *8.6 Secondary: prompt-only output behavior*

A separate study of 1,350 trials across nine configurations—900 adversarial, 20 tasks by five repetitions by nine cells, 895 counted because another model answered one Fable task—measured what models emit *after* the protected value is in context. Exposure had occurred in every adversarial trial by construction, so a clean cell shows output discipline, not protection. **Refusal is not non-disclosure**: of the 52 of 895 adversarial trials that leaked a protected value into the final answer, **27 were refusals** that quoted the value while declining the request—one mechanism by which a duty declared to a model is not a duty enforced, even when the model behaves as intended—and the other 25 were compliance. Leaks concentrate in a few cells and three of the nine configurations never leak, so the model is a nuisance variable. These figures are not comparable with the invocation rates of [17], and the configurations, whose system roles, tool surfaces and turn caps differ, are not a ranking.

## 9. Related Work

### *9.1 Classical authorization and policy administration*

Is the authorship hazard administrative access control under another name? The safety question was shown undecidable in the general protection model [23], and later models bought decidability with structure: roles mediating subjects and permissions [32], and administrative scopes governing who may change assignments [30]. Our approval stage is an instance of separation of duty [33]. Concurrent work applies separation of duty to an agent's privileged actions: Jing's governed execution architecture [34] interposes a planner, a policy gate, an executor and an auditor between the agent and the operating system, and binds each human approval to a digest of the staged bytes, consumable once. Our approval stage binds an approval the same way but to a different object, the governance artifact an agent drafts; [34] seals its policy file outside the agent's reach and leaves policy evolution to future work.

**The first divergence is where the untrusted party sits.** Administrative models protect authorization state by denying administrative roles; our agent lacks exactly that authority (Section 5.2), and the hazard survives because the agent writes the artifact an authorized party adopts. ARBAC97 has no vocabulary for a subject without administrative permissions whose output reaches the administrative decision as its content, which is why Table 1 marks the approver *trusted but bounded*.

**The second divergence is what counts as authorization state.** Administrative scoping governs role and permission assignment, not the vocabulary a policy's conditions dereference. Mandatory models do govern it: Bell–LaPadula's tranquility assumption forbids changing a label in use [35], declassification is treated as the security-relevant act [36], and attribute-administration models place attribute assignment under administrative control [37]. Moving `consignee` from F2 to F0 is a declassification, and Section 5.4 restores an MLS principle the dataspace agent setting has dropped, where labels are data and the party proposing a relabeling is the agent the label constrains. Tools built on administrative models miss it too: change-impact analysis [31] diffs two policy versions and security analysis for administrative RBAC [38] explores administrative operations, but a classification-only edit submits an identical policy. Our contribution is the observation that the administrative boundary must enclose the classification, with a check of what it costs when it does not.

**Software supply chains own the drafting half, for code.** An untrusted author whose change an honest reviewer merges is the attack pattern of [25], answered by two-person review [39], the control Section 5.3 places on the elevated path. The best-known attempt to measure whether reviewers catch adversarial changes used maintainers who had not consented and was withdrawn by its authors [40]; a measurement needs consenting reviewers. What supply chains do not cover is the object and the author: code means what its text says, a policy's meaning depends on a classification outside its text, and its drafter is of the class the policy binds.

Usage control [9] is the other inheritance: an ODRL duty enforced after access is a usage-control duty, and what we add is provenance and placement. The design's limits were named by the same literature: complete mediation [27] is what Section 6.1 applies, and a parent relaying a value it lawfully holds is a flow, not an access [21].

**Table 12.** Claim against evidence. The first row is the paper's title.

| Claim | Threats | Evidence |
|---|---|---|
| Authorship hazard; publication is a governance event (1.1) | — | Inspection (9.5), argument, and Case D: 75 of 90 preregistered single edits change authorization or duties when published ungated (8.4) |
| Classification is policy-rank authorship (5.4) | T2 | Case D: all six classification-only edits change authorization and narrow a duty with no policy text changed; the syntactic classifier passes all six and the amendment flags all six, both by construction; read as worded, the six conditions also pass 33 of 69 effective policy-text edits (8.4). Ablation of our own design, not a prior-work baseline; measures routing, not review. Registry-held classification (5.4) designed, not implemented in the prototype—Case D's evaluator models it |
| C1: agent as subject, authorization channel closed (5.1–5.2) | T1 | **Construction argument for (i)–(iii). Condition (iv) measured: 12 negotiations on Tractus-X EDC 0.13.0, every finalized agreement equal to the provider's stored policy; the bound is the stored definition, not the offer seen (8.3). No experiment attacks (i)–(iii)** |
| Per-action evaluation runs in the existing decision point (5.1, 7) | — | **Not yet. The stock validator rejects the Case D vocabulary and the policy contexts carry no field, purpose or depth; Case D's evaluator and Case A's decision client are reimplementations outside EDC (8.5, 10)** |
| Approval load is bounded per approver and floods are localized; central capacity scales with participants (5.3c) | T8 | **Case D: quota localizes flooding; central capacity is a provisioning parameter — 3 approvers hold to 200 participants and have diverged by 500 at the lowest draft rate, 30 hold to 3,000. Case D5: a federated partition keeps every local queue empty and moves the central limit to $K_c C/(\lambda\mu)$; μ unmeasured (8.4)** |
| The adopted classifier's review load is affordable (5.3b) | T8 | **Not measured. The review shares of Table 10 and the queue come from the earlier corpus under the worded conditions; D2 contains no benign edits (8.4)** |
| Profile exfiltration through the draft is rejected at submission (5.3b, 7) | T7 | **Bounded. Case D2 rejects verbatim and case- or space-altered sensitive values in every slot; a hyphenated variant passes in 2 of 5 slots; encoding, splitting and the choice of mapping are untested (8.4)** |
| C2 downward: duties act on values at invocation (6.1) | T3 | Case B, 105 cases and 1,320 calls (8.1, 8.2) |
| … for protected values not in named fields | T3 | **Contradicted. 7 of 7 exposed on the stress group (8.1)** |
| The constraint derives from the negotiated agreement (2.3) | — | Case E (8.3) |
| Per-action evaluation is affordable | — | **Not claimed. Case A measures the cost (8.5)** |
| Owner-declared ceiling on re-delegation (6.2) | T4 | **Design only; not claimed as a contribution** |

### *9.2 LLMs inside the enforcement path*

DAVE [1] is the closest system. Its post-generation layer is largely deterministic—pattern and entity detectors and similarity-based leakage control—with an optional LLM supervisor, and its probabilistic dependencies sit around the redactor: retrieval excludes chunks by NER-derived sensitivity tags, bounded by recognizer recall, and compliance requires span-level provenance of what the generator relied on, which no mechanism supplies. Our executor acts on fields whose class passed a publication gate (Sections 5.3–5.4); the difference is not that our classification is deterministic, since an agent drafts it too, but that it is governed. A 2026 literature improves DAVE's lane with policy-aware retrieval, redaction benchmarks and risk budgets across turns [41]; we reject an inferential component in the enforcement path rather than compete on its failure rate. In *model-mediated* enforcement a model or guard agent participates in the decision, as in GuardAgent [15]; in *deterministic architectural* enforcement decision and enforcement live outside the model, as in AgentSpec [14], Progent [13], the MCP proxy of [17] and our Case B. AgentSpec is closest on enforcement and differs on origin: its rules are authored for the runtime, ours derive from a negotiated agreement and must not be wider than it. ActGov [42], concurrent with this work, validates each proposed tool call outside the agent's reasoning, and its policies have a third origin: an LLM proposes and refines them from tool specifications, benign tasks and attack traces, an SMT check tests them against predefined invariants, and administrators confirm them, a step the paper names without specifying and drops in its fully automated variant. That is the structure of Section 1.1 outside dataspaces: the proposer is of the class its policies bind, in the automated variant the same model, and its inputs include the injected text of the attack traces.

### *9.3 Credentialing and capability*

The IDSA position paper [2] is the standards-track statement of the problem and anticipates both of our cases conceptually. The Rulebook's chapter on AI agents [8], unchanged from release 2026-1 to 2026-2, is descriptive: agents are issued tokens to execute negotiated contracts; in one pattern a consumer's MCP client "guarantees that policies of the data sharing contract will be respected"; and AI is recommended for mapping semantic models and for automated policy evaluation. It does not say who authorizes a mapping an agent produces, which is the question of Section 5.4; C2 is one way to make the client's guarantee hold, by enforcing the duty at the tool boundary (Section 6.1). Agent authority via verifiable credentials is otherwise pursued in OAuth, OIDC, DID and zero-knowledge settings rather than ODRL, and the Gaia-X ODRL credential profile covers organizations only. Agent infrastructure such as [29] propagates constraints across MCP and A2A chains with capability-based control inspired by Macaroons [28] and UCAN, and excludes post-access duties, usage control and runtime enforcement. Overlaying Governance [43] evaluates delegated agent authority from a relationship graph at each check rather than from a token, with scoped, depth-bounded and time-limited delegation in a single trust domain; any human who holds the permission roots a chain and the delegating agent issues onward delegations, so the data owner has no say distinct from the delegator's, which is the axis Section 6.2 adds.

### *9.4 Policy generation and enforcement*

AgentODRL [3] and ontology-guided instruction-to-ODRL translation [4] generate policies without governing their publication. ODRE [44] supplies an enforcement algorithm for ODRL and is the lineage we work in, and relationship-based authorization over an ODRL dataspace profile [45] compiles policy into a deterministic engine for organizations. Deterministic duty enforcement is mature in the IDS lineage —MYDATA and LUCON [46], IDS-RAM usage control [47] —implementing the usage-control model of [9]. Our executor is an implementation detail within that tradition; what is ours is that the obligation is carried from a negotiated agreement (Section 8.3) to the model's tool boundary without either end being restated by hand. AgenticRei [48] evaluates Rei deontic policies outside the model for tool calls and outbound A2A messages, surfaces obligations to the agent and monitors their discharge by deadline and audit, and notes that ODRL duties have “no enforcement model or lifecycle management”; Section 6.1 gives an enforcement model for duties over field classes, not a lifecycle. Its policies and ontologies are written by human authors, and it presents “extending the domain ontology” as “the policy update”, a benefit there and the risk of Section 5.4 here.

### *9.5 Agentic construction of governance artifacts: a targeted inspection*

This is the evidence behind Section 1.1. **Method.** We selected systems that use LLM agents to produce artifacts a dataspace would govern—ODRL policies, ontologies, schema alignments—from the venues and surveys cited there, read each evaluation section, and asked whether any step, human or automated, authorizes the generated artifact for use in a governed or production setting, as distinct from scoring its quality (Table 13).

**Table 13.** Targeted inspection of agentic generation systems for a publication or authorization step.

| System | Artifact | Quality evaluation | Depth | Auth. step |
|---|---|---|---|---|
| AgentODRL [3] | ODRL policy | 770 use cases; syntactic validity; SHACL loop | Full text | **None** |
| Mustafa et al. [4] | ODRL policy | Accuracy vs ontology-guided gold (91.95%) | Full text | **None** |
| Agent-OM [5] | Ontology alignment | OAEI tracks; precision/recall | Abstract + eval. | **None** |
| Talukder et al. [6] | Ontology | LLM-judge panel; competency-question SPARQL | Abstract + eval. | **None** |

The inspection covers four systems and is not exhaustive; two further systems from the same workshop venue were examined at abstract level only and omitted. It is not a criticism: a generator is evaluated on what it generates, and authorizing its output belongs to a literature that does not model the generator; in the systems inspected the seam between them is empty. The exception we found lies outside that scope: ActGov [42] names an administrator confirmation of LLM-proposed policies (Section 9.2). A negative result: the agent-driven semantic mapping that motivated this work is not defensible as a contribution against [5], on whose alignment quality we would lose.

## 10. Limitations

**The evidence is uneven** (Table 12). Per-action evaluation inside the connector's own decision point, the review cost of the adopted classifier and the owner-declared ceiling are not met or not measured; T7 is met for verbatim values only; the closure of the authorization channel is measured at one of its four conditions, with no experiment attacking (i) to (iii); and two filled rows go against the design. The classification row carries the sharpest result. The limitations follow in order of how much they should worry a reader.

**Composition (T5) is unaddressed.** A planner can decompose a forbidden goal into individually permitted actions across itself and its sub-agents, and a stateless per-action evaluator admits every one; the advertised properties, no token and no persistent state, make the design blind here. A session-scoped budget over sensitive-field exposure, in the spirit of [41], evaluated by the PDP on state the agreement owns, is the mitigation we consider; it is not designed. The incident report draws the general lesson that agents “combine vulnerabilities, credentials, and permissions into attack paths that may not be apparent when individual weaknesses are assessed separately” [18]; T5 is that lesson restricted to permitted actions.

**The ceiling is an access boundary, not a confinement boundary (T6).** A parent can relay a field it lawfully holds to a child in a prompt, and no evaluation at the PDP sees that flow [21]; the design establishes no non-interference property, which [1] also declines to claim. Reaching T6 needs a gateway on the A2A channel applying the child's ceiling to the parent's outbound messages, or taint propagation through the planner. AgenticRei [48] intercepts outbound A2A messages with the engine it uses for tool calls, which shows that such a check point is feasible; it evaluates the sender's action, not the receiver's ceiling. The incident of Section 2.3 shows the relay in practice: agents exchanged credentials and techniques through a message board they built in the package mirror, using the credentials they had been issued to install packages and exploiting no vulnerability to do so [18].

**The guarantees assume that trusted components hold.** Section 4 places their compromise out of scope, but an agent assumed fully adversarial will probe them. In the incident of Section 2.3, agents used the shared credential they had been issued and a token-refresh flaw in the package mirror to turn a forged administrative token into a valid one, created an administrator account and installed plugins, and later stole the mirror's token-signing key and forged administrator tokens with it [18]; a stolen connector key would defeat the lineage chain of Section 6.2.3 in the same way. A single decision authority does not preclude the overlapping, independent controls the report calls for [18], but the design has one enforcement point per objective (Table 5) and does not duplicate them; the gateway is deterministic and narrow to keep its surface small, and its hardening is not evaluated.

**Value-level enforcement holds for named fields only** (Section 8.1). "Anonymize" means suppression: `AnonymizeFieldClass`, named after the stock `Anonymize` action it attaches to, removes a class's fields from the model-bound payload and does not anonymize a record statistically. In the long tail of Section 3 the permitted F0 and F1 fields may identify a small exporter by themselves, so derivable cases may be the rule there; we have not measured how often.

**The influence channel is bounded, not closed.** Approver fatigue and social engineering remain, and Section 5.4 moves reclassification onto the elevated path without removing the approver. The reviewer's hit rate on a reclassification that reads as a label correction is the unmeasured quantity on which the routing results of Table 8 depend, and the review cost of the wider classifier is unmeasured, since Table 10 rests on the earlier corpus under the worded conditions.

**Central approval capacity scales with participants.** Nothing forces a single central pool: the registration agent runs inside the participant boundary and the profile never leaves it, so the participant is the party best placed, and already responsible under the sovereignty model, to review its own asset's mapping and policy. Under federated approval only changes to classification, the registry and the classifier need central review, and Case D5 shows the exit exists but is narrow at the earlier corpus's central-rank share: a federated pool of three holds 292 participants against the centralized 100 to 250. Whether central-rank changes are rare is a deployment measurement.

**Per-action evaluation does not yet run in EDC's decision point.** Stock Tractus-X EDC 0.13.0 refuses the Case D vocabulary (ten registrations, all rejected), and its policy contexts carry the participant, the agreement and the time but no field, purpose or delegation depth. Supplying them means registering constraint functions, as in Appendix A, which would re-implement the evaluator inside EDC; the decision authority stays single, but the engine is not used as is. The agreement supplies the duty at class level (anonymize F2), while the mapping from class to field names is local metadata in this prototype, the artifact Section 5.4 places under the gate, so the *normative rule* comes from the agreement, not every parameter. Public keys are resolved from a fixed test key pool, while signatures are verified and expired, revoked, wrong-subject and missing-credential controls fail. The connector does not redact; the gateway does. Scope is one tool and a small action set on an upstream test harness.

**There is no external baseline.** No benchmark exists for governed agentic data access, and DAVE [1] does not yet implement or empirically evaluate its full pipeline, so prompt-only, discovery filtering and Case D's no-gate condition are our constructions. Change-impact analysis [31] over the D2 edit space, which a classification-only edit defeats by construction (Section 9.1), would need Margrave and an XACML decision point against which a translation of our ODRL dialect could be validated; we have neither. Releasing corpus, scorer and raw trials improves reproducibility, not external validity.

**Case C was not run**, and it would test Section 6.2: *C-utility* asks how often the default ceiling blocks a decomposition a domain expert judges legitimate, and *C-composition* whether an adversarial planner can combine permitted actions into a violation (T5), which we expect to succeed in some cases. Where branch-level revocation or offline operation is needed, a capability design is better (Section 6.2.4). Case A as originally scoped also asked whether a revoked agreement denies at the next action; we took that to hold by construction and measured latency and revocation timing instead, and Section 8.5 shows that it holds only when the decision path queries the revocation register, which the agreement object does not reflect. **Case E rests on a profile extension we wrote**, small but not standardized, on a base profile that is itself an unofficial draft. The claim is feasibility, not adoption: eighteen non-blank lines in four existing files plus a 63-line constraint function and a 35-line schema, with validation left on and the rest of the path unmodified—not that the term is already available. **Conceptual priority on agent authority is [2]'s.** The IDSA Rulebook already carries a chapter on AI agents, so far without normative requirements [8], and a task force is active; should a normative specification land first, this work becomes a comparison against it.

## 11. Conclusion

When agents enter a governed environment the temptation is to add a model to the enforcement path or a verification stack beside the policy plane. We argued instead that the literature in which agents *write* governance artifacts never models publication, and that once publication is seen as the governance event, the hazard of an agent authoring its own norms, and the quieter one of authoring the classification they refer to, become visible. Closing the authorization channel is structural and cheap; bounding the influence channel is enforcement and ongoing. Leaving authority in the agreement mints no rights-bearing artifact, at the price of coarse revocation and coupling to the decision point. The downward mechanism holds on the easy half of its problem and fails, seven cases of seven, on the other; the constraint traces to a contract two parties negotiated. The hazard's mechanism is verified over a 344,512-request universe, where every classification-only edit changes decisions that a policy-diff classifier cannot see. A central reviewer pool must be staffed at one reviewer per 20 to 138 participants, and federation moves that limit onto a share we did not measure. Whether reviewers catch what is flagged remains the open human-factors question.

## Author Contributions

S.L. conceived the authorship-hazard framing and Principle P, designed C1 and C2 including the approval stage, the owner-declared ceiling and the attestation chain, and wrote the paper. C.L. designed and executed every experiment in Section 8—the harness, corpora and stress group of Cases B and E, the clean-environment rerun of Case B's live-model studies, the duty compiler and gateway, the negotiation experiment and profile extension, the preregistered Case D2 with its evaluator, classifiers, edit generator and request universe, the earlier Case D corpus with its queue simulation and Case D5, and the testbed of Case A—including the two experiments that produce results against the design. Both authors agreed the claims and limitations.

## Appendix A. The profile extension

The obligation term added to the Catena-X policy profile in Section 8.3. Its meaning, carried in the schema as an annotation: *before data covered by this agreement is released into an agent execution context, fields classified with the sensitivity class specified herein shall be suppressed.* It is a single atomic constraint, `cx-policy:AnonymizeFieldClass eq "F2"`, with leftOperand fixed to `AnonymizeFieldClass`, operator `eq`, rightOperand matching `^F[0-9]$`, and no additional properties. The change adds eighteen non-blank lines to four existing files, removes one, and adds a 63-line constraint function and a 35-line schema; the 82-line diff (SHA-256 `d5bad0bb…`) is in the evidence package. Validation is not disabled anywhere, and an existing stock term still registers and negotiates correctly, which is the regression control.

| File | Change |
|---|---|
| Policy validation constants | One literal added; one entry appended to the allowed obligation left-operand list |
| Atomic constraint schema index | One `$ref` to the new schema |
| Policy extension | Constraint function registered for the duty rule type in the transfer-process and policy-monitor contexts |
| JSON-LD context document | One term definition binding the label to its profile IRI |
| **New: constraint function** | Evaluates the duty constraint |
| **New: constraint schema** | The shape above, with the obligation text as an annotation |

## Appendix B. Changes from the first version

(1) Case D is reported as the preregistered corrective replication D2, because an audit of the first version's constructed validation found repeated edits, constraint widenings that widened nothing, no prohibition in the base state, and an evaluator that read neither assignees nor duty targets. (2) The privilege-delta conditions are stated in their wider reading. (3) Table 11 gives the agreement-only control its own column; the first version placed it in the admission-time column. Decisions permitted after revocation are 99 at a five-second cache, 591 at thirty seconds and 439 on the agreement-only path, not 101, 593 and 440, because the first version's counts also included the admission decision and, where there was one, the final denial. (4) Sections 8.2 and 8.6 report a preregistered rerun of every live-model configuration in a clean environment. The first version's Claude trials carried no turn between trials but did receive runtime-level context that the settings flags did not exclude: an automatic memory and the account e-mail in Section 8.2, and the account e-mail and working path in Section 8.6. One Section 8.6 cell, whose runtime home cannot be rebuilt, is replaced by the same runtime version in a clean home. The rerun counts a trial only when the requested model answered it, which excludes one Fable task, and changes the figures: prompt-only and discovery filtering leak 56 and 54 of 219 adversarial trials, not 49 and 57 of 220, with benign success 219 of 220 in each condition; three prompt-only cells are clean by output, not five; and Section 8.6 counts 52 of 895 adversarial leaks, 27 of them refusals, not 39 of 900 and 24. Every preregistered claim holds. (5) Section 8.4 counts the central-rank share as 130 of 380, not the 80 reclassifications alone, since the federated partition also routes vocabulary and classifier edits centrally; it replaces a grid-resolution tie between the federated and centralized pools with a closed-form comparison, and separates the amendment's simulated 1.83 maximum from the 1.93 limit. (6) The profile extension includes two new files. (7) T7 is bounded for verbatim values only and Table 12 (Table 13 in the first version) gains three rows. (8) Sections 2.3 and 10 cite the July 2026 OpenAI–Hugging Face evaluation incident [18] for the execution boundary and for the limits on trusted components, composition and relay. (9) Sections 1.1, 9.3 and 10 cite the chapter on AI agents that the IDSA Rulebook has carried since release 2026-1 [8], which the first version described as pending. (10) Section 9 discusses four 2026 systems on agent governance [34, 42, 43, 48]. The paper was also shortened, and the former subsection of Section 8 on what the measurements do not show is merged into Section 10.

**Evidence package.** Preregistrations for Case B's extension (`3484d360…`), its clean-environment live-model runs (CR-1, `d5b5760b…`), Case D5 (`94b52959…`) and Case D2 (`20d86cf9…`) were frozen by combined SHA-256 before their corpora or runs; CR-1's post-freeze changes are kept as timestamped amendments. Raw records, corpora and recomputation scripts carry a hash for every file; the package is at huggingface.co/datasets/ktnet-research/subjects-not-authors-evidence (doi:10.57967/hf/10777).

## References

1. R. Brinkhege and P. Menon. DAVE: A Policy-Enforcing LLM Spokesperson for Secure Multi-Document Data Sharing. arXiv:2602.17413, Feb. 2026.
2. International Data Spaces Association. Data Spaces and AI: Trustworthy Agentic Participation in Data Spaces. Position paper, version 1.0, 9 July 2026. DOI 10.5281/zenodo.21279055.
3. W. Zhong, K. Huang and X. Du. AgentODRL: A Large Language Model-based Multi-agent System for ODRL Generation. Accepted at AAAI 2026. arXiv:2512.00602, Nov. 2025.
4. D. M. Mustafa, A. Nadgeri, D. Collarana, B. T. Arnold, C. Quix, C. Lange and S. Decker. From Instructions to ODRL Usage Policies: An Ontology Guided Approach. LLM+KG Workshop at VLDB 2024, Guangzhou, Aug. 2024; posted as arXiv:2506.03301, June 2025.
5. Z. Qiang, W. Wang and K. Taylor. Agent-OM: Leveraging LLM Agents for Ontology Matching. *Proceedings of the VLDB Endowment*, 18(3):516–529, 2024. DOI 10.14778/3712221.3712222.
6. A. Talukder, M. A. Mridul and O. Seneviratne. Towards Automated Ontology Generation from Unstructured Text: A Multi-Agent LLM Approach. arXiv:2604.23090, Apr. 2026.
7. H. Bian. LLM-Empowered Knowledge Graph Construction: A Survey. arXiv:2510.20345, Oct. 2025.
8. International Data Spaces Association. IDSA Rulebook, release 2026-2, chapter “AI and Dataspaces: Shaping the Future of Trusted, Decentralized Intelligence” (documentation/130_AI_Agents.md; text unchanged since release 2026-1 of 27 Apr. 2026). 29 Jul. 2026. https://github.com/International-Data-Spaces-Association/IDSA-Rulebook, retrieved 1 Oct. 2026.
9. J. Park and R. Sandhu. The $UCON_{ABC}$ Usage Control Model. *ACM Transactions on Information and System Security*, 7(1):128–174, 2004. DOI 10.1145/984334.984339.
10. Salvachúa, Muñoz, Huecas, Aparicio and Menendez (UPM/IPTC). ODRL Profile: Data Spaces. Unofficial draft in the W3C ODRL Community Group repository, undated. https://w3c.github.io/odrl/profile-dataspaces/, retrieved 22 Sep. 2026.
11. Q. Zhan, Z. Liang, Z. Ying and D. Kang. InjecAgent: Benchmarking Indirect Prompt Injections in Tool-Integrated Large Language Model Agents. In *Findings of the Association for Computational Linguistics: ACL 2024*, pp. 10471–10506, 2024. DOI 10.18653/v1/2024.findings-acl.624.
12. E. Wallace, K. Xiao, R. Leike, L. Weng, J. Heidecke and A. Beutel. The Instruction Hierarchy: Training LLMs to Prioritize Privileged Instructions. arXiv:2404.13208, 2024.
13. T. Shi, J. He, Z. Wang, H. Li, L. Wu, W. Guo and D. Song. Progent: Securing AI Agents with Privilege Control. arXiv:2504.11703, 2025.
14. H. Wang, C. M. Poskitt and J. Sun. AgentSpec: Customizable Runtime Enforcement for Safe and Reliable LLM Agents. In *Proc. 48th IEEE/ACM International Conference on Software Engineering (ICSE)*, 2026.

15. Z. Xiang et al. GuardAgent: Safeguard LLM Agents by a Guard Agent via Knowledge-Enabled Reasoning. In *Proc. 42nd International Conference on Machine Learning (ICML)*, 2025.
16. J. Lichtefeld, J. A. Cecil, A. Hedges, J. Abramson and M. Freedman. Redacted Contextual Question Answering with Generative Large Language Models. In *Proc. First Int. Conf. on NLP and AI for Cyber Security (NLPAICS)*, pp. 230–237, Lancaster, UK, 2024.
17. R. Uppala. Prompts Don't Protect: Architectural Enforcement via MCP Proxy for LLM Tool Access Control. arXiv:2605.18414, May 2026.
18. OpenAI. OpenAI – Hugging Face Incident Technical Report. 26 Aug. 2026. https://cdn.openai.com/pdf/67869394-cb91-4c12-888c-5cbd85c7814c/OpenAI-Hugging-Face%20Incident-Technical-Report.pdf, retrieved 1 Oct. 2026.
19. M. Franklin, A. Halevy and D. Maier. From Databases to Dataspaces: A New Abstraction for Information Management. *SIGMOD Record*, 34(4), 2005.
20. R. G. Smith. The Contract Net Protocol: High-Level Communication and Control in a Distributed Problem Solver. *IEEE Transactions on Computers*, C-29(12), 1980.
21. D. E. Denning. A Lattice Model of Secure Information Flow. *Communications of the ACM*, 19(5):236–243, 1976. DOI 10.1145/360051.360056.
22. A. J. I. Jones and M. Sergot. On the Characterisation of Law and Computer Systems: The Normative Systems Perspective. In *Deontic Logic in Computer Science*, 1993.
23. M. A. Harrison, W. L. Ruzzo and J. D. Ullman. Protection in Operating Systems. *Communications of the ACM*, 19(8):461–471, 1976.
24. N. Hardy. The Confused Deputy (or Why Capabilities Might Have Been Invented). *ACM SIGOPS Operating Systems Review*, 22(4):36–38, 1988. DOI 10.1145/54289.871709.
25. P. Ladisa, H. Plate, M. Martinez and O. Barais. SoK: Taxonomy of Attacks on Open-Source Software Supply Chains. In *Proc. IEEE Symposium on Security and Privacy (SP)*, pp. 1509–1526, 2023.
26. N. Boucher and R. Anderson. Trojan Source: Invisible Vulnerabilities. In *Proc. 32nd USENIX Security Symposium*, pp. 6507–6524, 2023.
27. J. H. Saltzer and M. D. Schroeder. The Protection of Information in Computer Systems. *Proceedings of the IEEE*, 63(9):1278–1308, 1975. DOI 10.1109/PROC.1975.9939.
28. A. Birgisson, J. G. Politz, Ú. Erlingsson, A. Taly, M. Vrable and M. Lentczner. Macaroons: Cookies with Contextual Caveats for Decentralized Authorization in the Cloud. *NDSS*, 2014.
29. S. Prakash. AIP: Agent Identity Protocol for Verifiable Delegation Across MCP and A2A. arXiv:2603.24775, Mar. 2026.
30. R. Sandhu, V. Bhamidipati and Q. Munawer. The ARBAC97 Model for Role-Based Administration of Roles. *ACM Transactions on Information and System Security*, 2(1):105–135, 1999. DOI 10.1145/300830.300839.
31. K. Fisler, S. Krishnamurthi, L. A. Meyerovich and M. C. Tschantz. Verification and Change-Impact Analysis of Access-Control Policies. In *Proc. 27th International Conference on Software Engineering (ICSE)*, pp. 196–205, 2005. DOI 10.1145/1062455.1062502.
32. R. S. Sandhu, E. J. Coyne, H. L. Feinstein and C. E. Youman. Role-Based Access Control Models. *IEEE Computer*, 29(2):38–47, 1996. DOI 10.1109/2.485845.
33. D. D. Clark and D. R. Wilson. A Comparison of Commercial and Military Computer Security Policies. In *Proc. IEEE Symposium on Security and Privacy*, pp. 184–194, 1987.
34. Q. Jing. Separation of Duties for Privileged LLM Agents: A Governed Execution Architecture with Measured Security-Utility Trade-offs. arXiv:2609.38224, Sep. 2026.
35. D. E. Bell and L. J. LaPadula. Secure Computer System: Unified Exposition and Multics Interpretation. Technical Report MTR-2997 Rev. 1, The MITRE Corporation, Bedford, MA, Mar. 1976.
36. A. Sabelfeld and D. Sands. Declassification: Dimensions and Principles. *Journal of Computer Security*, 17(5):517–548, 2009. DOI 10.3233/JCS-2009-0352.
37. X. Jin, R. Krishnan and R. Sandhu. A Role-Based Administration Model for Attributes. In *Proc. First International Workshop on Secure and Resilient Architectures and Systems (SRAS)*, 2012. DOI 10.1145/2420936.2420938.
38. N. Li and M. V. Tripunitara. Security Analysis in Role-Based Access Control. *ACM Transactions on Information and System Security*, 9(4):391–420, 2006. DOI 10.1145/1187441.1187442.
39. The SLSA Project (OpenSSF). SLSA: Supply-chain Levels for Software Artifacts, version 0.1, Requirements, “Two-person reviewed” (SLSA 4). 2021. https://slsa.dev/spec/v0.1/requirements
40. IEEE S&P 2021 Program Committee. IEEE S&P’21 Program Committee Statement Regarding the “Hypocrite Commits” Paper. 6 May 2021. https://www.ieee-security.org/TC/SP2021/downloads/2021_PC_Statement.pdf
41. Q. Yang, Y. Li and X. Ma. Private-RAG: A Privacy-Preserving Retrieval-Augmented Generation Method for Large Model Inference. *Electronics*, 15(12):2567, June 2026. DOI 10.3390/electronics15122567.
42. K. Zhang, Y. Peng, K. Jiang and Y. Zhang. ActGov: Governing LLM Agent Actions via Policy-Constrained Validation. arXiv:2609.24446, Sep. 2026.
43. A. Ibrahim and Y. Li. Overlaying Governance: A Compositional Authorization Framework for Delegation and Scope in Agentic AI. arXiv:2606.03518, June 2026.
44. A. Cimmino, J. Cano-Benito and R. García-Castro. Open Digital Rights Enforcement Framework (ODRE): From Descriptive to Enforceable Policies. *Computers & Security*, 150:104282, 2025. DOI 10.1016/j.cose.2024.104282.
45. I. Plaza-Ortiz et al. Authentication and Authorization in Data Spaces: A Relationship-Based Access Control Approach Based on ODRL. OPAL 2025. arXiv:2505.24742.
46. International Data Spaces Association. Data Usage Control Technologies (MYDATA, LUCON). Web page, 11 Sep. 2020. https://internationaldataspaces.org/data-usage-control-technologies/
47. International Data Spaces Association. IDS Reference Architecture Model 4 (IDS-RAM 4), §4.1.6 Usage Control. 2022. https://docs.internationaldataspaces.org/ids-knowledgebase/ids-ram-4/
48. A. Joshi, T. Finin, K. P. Joshi and L. Kagal. Deontic Policies for Runtime Governance of Agentic AI Systems. In *IEEE Symposium on Agentic Services (IEEE ICWS 2026)*, to appear; arXiv:2606.19464, June 2026.

Prior-art claims were checked against primary sources on 18 September 2026, and those added in this version on 1 October 2026. All figures in Section 8 were recomputed from the raw trial records rather than transcribed from run summaries.